# Contra-Intuitive Features of Time-Domain Brillouin Scattering in Collinear Paraxial Sound and Light Beams

Vitalyi E. Gusev

Laboratoire d'Acoustique de l'Université du Mans, LAUM - UMR 6613 CNRS, Le Mans Université, Avenue Olivier Messiaen, 72085 Le Mans cedex 9, France.

**Abstract**

Time-domain Brillouin scattering is an opto-acousto-optical probe technique for the evaluation of the transparent materials. Ultrashort pump laser pulses via optoacoustic conversion launch in the sample picosecond coherent acoustic pulses. The time-delayed ultrashort probe laser pulses monitor the propagation of the coherent acoustic pulses via photo-elastic effect, which induces light scattering. A photodetector collects acoustically scattered light and the probe light reflected by the sample structure for the heterodyning. The scattered probe light carriers the information on the acoustical, optical and acousto-optical parameters of the material in the current position of the coherent acoustic pulse. Thus, among other applications, the time-domain Brillouin scattering is a technique for three-dimensional imaging. Sharp focusing of the coherent acoustic pulses and probe laser pulses could increase lateral spatial resolution of imaging, but could potentially diminish the depth of imaging. However, the theoretical analysis presented in this manuscript contra-intuitively demonstrates that the depth and spectral resolution of the time-domain Brillouin scattering imaging, with collinearly propagating paraxial sound and light beams, do not depend at all on the focusing/diffraction of sound. The variations of the amplitude of the time-domain Brillouin scattering signal are only due to the variations of the probe light amplitude caused by light focusing/diffraction. Although the amplitude of the acoustically scattered light is proportional to the product of the local acoustical and probe light field amplitudes the temporal dynamics of the time-domain Brillouin scattering signal amplitude is independent of the dynamics of the coherent acoustic pulse amplitude.

**Keywords:** photoacoustics, optoacoustics, laser ultrasound, time-domain Brillouin scattering, picosecond acoustic interferometry, imaging, microscopy

# 1. INTRODUCTION

The applications of the optical pump-probe technique with ultrashort laser pulses for the investigation of the coherent acoustic waves in the GHz frequency range are under the development for more than 35 years already [1-4]. To access high acoustic frequencies range the technique employs the laser pulses of picosecond to femtosecond duration. The absorption of the pump laser pulses launches inside the media the coherent acoustic pulses (CAPs), which are of picosecond duration or contain the fronts of picosecond duration [1-3,5]. Consequently, the spectrum of the CAPs contains GHz acoustic frequencies efficient in the Brillouin scattering of light [6,7]. Thus, the propagation of the CAPs in transparent media can be followed via the acoustically induced scattering of the time-delayed probe laser pulses. The amplitude of the scattered light is proportional to the CAP amplitude and is commonly much smaller than the amplitude of the probe light reflected at the sample surface and by other possible stationary features of the sample structure. Thus, an essential part of the technique is optical heterodyning, achieved by directing the probe light scattered by the CAPs and the probe light reflected by the sample surfaces/interfaces to a common photodetector [1-3]. The photodetector monitors the interference of the weak light scattered by the nearly transparent moving acoustical mirror and the light reflected by the stationary optical inhomogeneities of the sample. As the delay between the phases of the scattered and the reflected light changes in time due to CAP propagation, the signal of the photodetector contains an oscillating part replicating in time the transition from constructive to destructive interference when the CAP penetrates deeper and deeper inside the medium. This ultrafast pump-probe technique was initially called picosecond acoustic interferometry [2,3,8]. Recently the name time-domain Brillouin scattering (TDBS), which highlights the fundamental physical phenomenon involved and the difference from the classical frequency-domain Brillouin scattering [6,7] has become more common [9]. In fact, the frequency of the coherent oscillation revealed by the TDBS technique in the homogeneous medium is equal to the frequency shift of the Stokes and anti-Stokes light measured by FDBS, i.e. to the Brillouin frequency (BF). The detected acoustically induced oscillation revealed by the photodetector is called the Brillouin oscillation (BO).

One of the fundamental applications of the TDBS is the evaluation of the absorption of the phonons in the GHz –THz frequency range [2,8,10-17]. This is achieved by measuring the decay in time of the BO amplitude. The pioneer research paper [8] contains the estimates of the other physical factors that could cause the attenuation of the BO and should be either excluded experimentally or properly taken into account to reveal acoustic absorption. Among them are the coherence length of the probe laser pulses, attenuation of the probe light in the medium and the diffraction of both CAPs and the probe laser pulses. Particularly, the result of the theoretical analysis presented in Ref. [8]

predicts, as it could be expected, that the characteristic spatial scale for the influence of the diffraction phenomena is the Rayleigh range [18,19] also sometimes called the diffraction length [20-22]. The Rayleigh range is proportional to the square of the wave beam radius at $1/e^2$ level and inverse proportional to the wavelength, $z_R \sim \pi a^2/\lambda$ . However, the theory is developed in Ref. [8] for the equal radii of the acoustic and light beams, while the acoustic wavelength in the considered backward Brillouin scattering is tightly related to the probe light wavelength in the medium, $\lambda_{acoustical} = \lambda_{probe}/2$ . Consequently, it is impossible to disentangle the roles of the acoustical and optical diffraction in the predicted decay of the BO amplitude. In the most of the subsequent research devoted to the evaluation of the acoustic absorption by the TDBS the role of the diffraction phenomena was excluded experimentally by avoiding sharp focusing of both the pump laser pulses, which controls the radius of the launched CAP beam [19], and of the probe laser pulses. The Rayleigh ranges of both the CAP and the probe light were significantly exceeding the acoustical absorption length.

The interest to the role of the diffraction phenomena in the TDBS has significantly increased quite recently mostly due to the development of three specific applications of the TDBS. The TDBS is applied to monitor the CAPs propagation in samples with the characteristic dimensions smaller than several micrometers, which requires tighter focusing of laser pulses than in the sound absorption measurements [15,23]. The TDBS technique is applied to monitor the CAPs emitted by the submicron and nanometer size objects (optoacoustic transducers) [15,24-26]. In this case the radius of the launched CAP beam is controlled not by the radius of the pump laser focus but by the dimensions of the laser-irradiated object, which could be much smaller in the case of the nanoparticles or nanowires, for example [27-29]. Finally, the TDBS is applied for the imaging in of the spatial inhomogeneities in acoustical, optical and acousto-optic parameters of the transparent media [9,27-29]. The depth spatial resolution of imaging is commonly sub-optical, because the acoustic wavelength in the TDBS is twice shorter than the probe optical wavelength in the medium [17,30,31]. Theoretically, it is limited by the axial localization of the light scattering acoustic mirror, i.e. by the length of the strain CAP or the width of the strain front in the CAP [9,24-26]. In application to three-dimensional imaging [17,30,31] the lateral spatial resolution is controlled by the focusing of the pump and probe light. A rather sharp focusing of laser pulses is required to make lateral resolution comparable to depth resolution. As the increased focusing leads to the diminishing of the Rayleigh ranges for the acoustic and probe light beams, the theoretical studies of the diffraction role in the TDBS are required to understand the trade-off between the increased spatial resolution and decreased depth of imaging. It is also worth mentioning that the decay in time of the TDBS signal amplitude, limiting the depth of imaging, broadens the spectral width of the BO, thus influencing the frequency resolution of the TDBS in possible spectroscopic application for materials identification. This provides additional motivation for the development of the analytical theory of the TDBS in diffracting sound and light beams.

The influence of the diffraction effects on the TDBS signals was reported in the experiments [15,32,33]. In all of them, the deviation in the temporal decay of the detected BO from an exponential one was revealed with increasing propagation distance of laser-generated CAP from the emitting opto-acoustic transducer. In Ref. [15] it was hypothesized that the revealed attenuation of the BO, additional to acoustical absorption, is caused by the diffraction of the coherent acoustic waves emitted by a sub-micrometer contact between the transparent sphere and laser-irradiated metallic surface. For the contact radius of 0.3 μm this hypothesis was supported by the estimates of the characteristic diffraction length of the coherent acoustic wave at BF. The experiments reported in Refs. [32] and [33] are directly addressing the trade-off between the lateral spatial resolution and the depth and the spectral resolution of the TDBS imaging by investigating the influence of the focusing of the pump and probe laser pulses on the TDBS signal decay. The radii of the pump and probe laser foci were modified using microscope objectives with the different numerical apertures (NA). Based on the estimates of the Rayleigh ranges, the decay in TDBS amplitude, additional to one caused by the acoustics absorption, was attributed in Ref. [32] mostly to the acoustic diffraction, while in Ref. [33] to the diffraction of both CAPs, $\sim 1/\sqrt{1 + \left[z_a(t)/z_R^{acoustical}\right]^2}$, and of the probe light, $\sim 1/\sqrt{1 + \left[z_a(t)/z_R^{probe}\right]^2}$. Here $z_a(t)$ is the distance of the emitted picosecond CAP from the common focus of the pump light and of the probe light on the optoacoustic transducer. $z_a(t)$ is the depth at which the Brillouin scattering takes place after the photo-generation of the CAP by the pump laser pulse at time t=0. The model [33] assumes that the TDBS signal amplitude is proportional to the product of the local amplitudes of the CAP and the probe light field, as it could be expected from the local law of the acousto-optic (photo-elastic) effect [3,4,6,7].

The analytical theory presented below provides opportunity to disentangle the roles of the coherent sound and light diffraction in the TDBS experiments conducted with collinear paraxial acoustical and probe optical beams. It contra-intuitively predicts that although the amplitude of the light wave scattered by the CAP is proportional to the product of the probe light amplitude and the CAP amplitude, the variation of the TDBS signal amplitude in time replicates the probe light spatial distribution and is completely independent of the variations with time of the CAP amplitude. Thus, the TDBS signal dynamics depends on the diffraction/focusing of the probe light and is independent of the diffraction/focusing dynamics of the coherent acoustic field. Unexpectedly, the diffraction/focusing of the paraxial CAP beam has no influence on the depth of imaging and the spectral resolution of the TDBS technique. The theory attributes the physical origin of the predicted phenomenon to the heterodyne detection, which shapes in the wave-vectors and frequencies spaces a specific sensitivity function for monitoring of the coherent acoustic beams by the Brillouin scattering. This theoretical prediction provides new insights in the interpretation of the earlier TDBS experiments [15,23,32,33] and in the possible optimization of the TDBS technique for nanoscale imaging and spectroscopic applications in bio-medical and material sciences.

The manuscript is structured as follows. In Section II, the theoretical analysis of the probe light beam TDBS by the CAP beam is presented, first, for the general case of arbitrary shaped paraxial co-focused beams. Second, it is specified for the case of the Gaussian beams. Third, the theoretical predictions for the Gaussian collinear paraxial acoustic and light beams with different spatial positions of their foci are described. In Section III, the theoretical predictions are compared with and are used for the interpretation of the known experimental results. A short comparison of the TDBS with the stimulated Brillouin scattering (SBS) [34,35] and the SBS-related phenomenon of the light wave front/phase conjugation [35-37] is also placed in Discussion section followed by the Conclusions.

## 2. THEORY

In the TDBS several different experimental configuration are possible. On the one hand, they differ by the position of the optoacoustic transducer for launching the CAPs and the internal mechanical/elastic structure of the sample, influencing the CAPs propagation (reflections/refractions). On the other hand, they differ by the optical structure of the sample, which is simultaneously creating an initial distribution of the optical field inside the sample for probing the CAPs and a reflected optical field for the heterodyning of the light scattered by the CAPs. However, all the experimental configurations based on the collinearly propagating CAPs and probe laser pulses are sharing the same background physical principles. In the presence of the CAPs the probe light waves initially launched in the sample are additionally backward scattered independently of their propagation direction relative to the CAPs propagation direction. This backward Brillouin scattering is accompanied by the emission or the absorption of the acoustic phonons and, respectively, by downshifting or upshifting of the scattered light spectral line. The scattered Stocks and/or anti-Stocks light and the reflected probe light are collected by a photodetector for the optical heterodyne detection. These measurements of the light total energy reveal the temporal oscillations at the scattered and reflected light difference frequency (BF), which is equal to the frequency of the acoustic phonon.

### 2.1. Arbitrary paraxial co-focused acoustic and light beams

A particular example of the experimental TDBS scheme presented in Fig. 1 comes from the seminal publication [8]. The theory presented below extends the analysis conducted in Ref. [8] for the collinear propagation of plane probe light and coherent acoustic pulses to the case of the collinear propagation of the diffracting CAP and probe light beams.

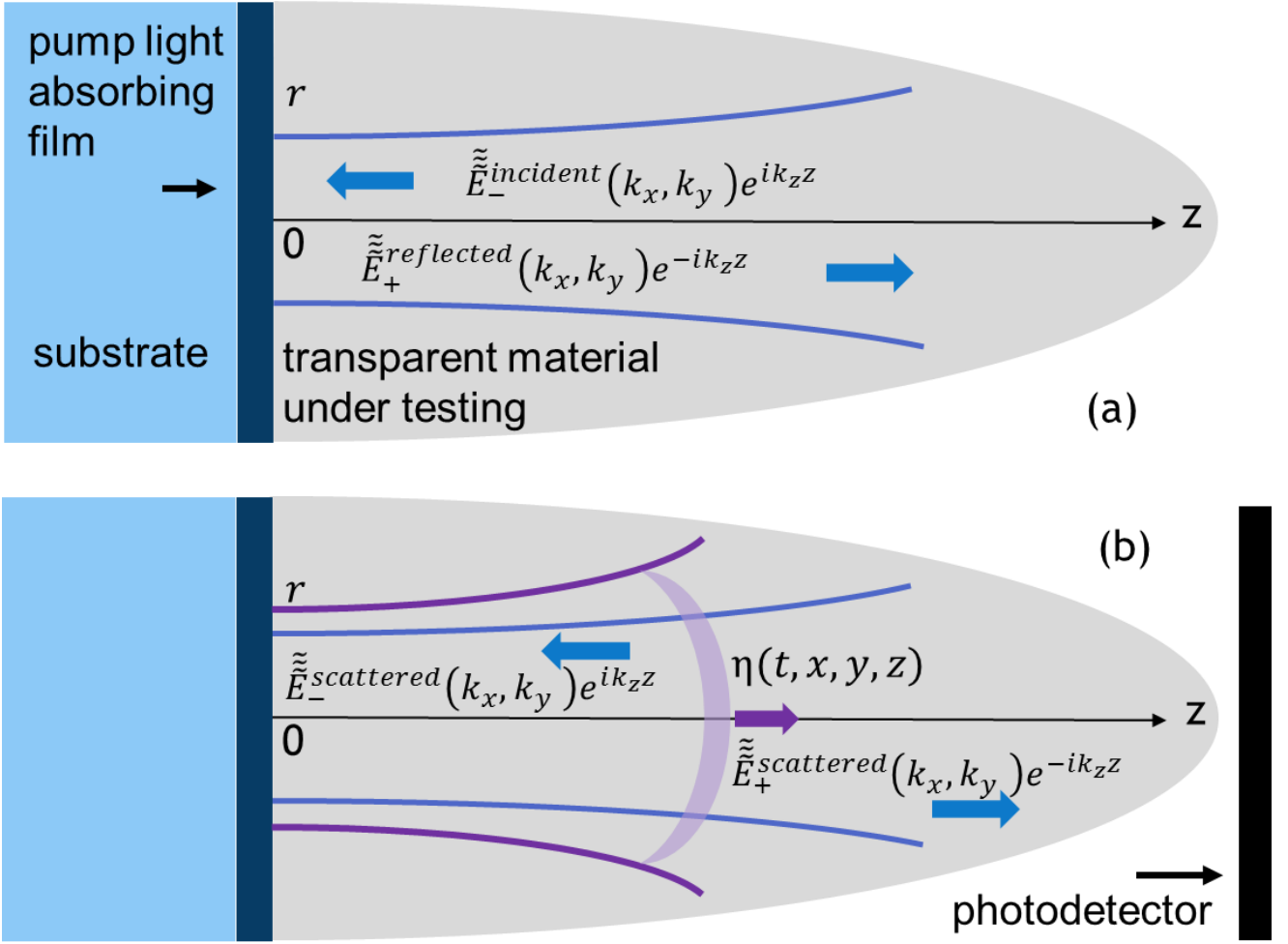

**Figure 1**. **Schematic presentation of co-focused probe light and coherent sound fields contributing to time-domain Brillouin scattering signal detection. a** Initial probe light field, that could be created in the system in the absence of the coherent acoustic field, is due to the laser pulses incident from the light transparent medium z>0 on the sample surface z=0 and the laser pulses reflected from this surface. **b** Coherent acoustic pulse $\eta(t,x,y,z)$ is launched in the transparent media z>0 due the absorption in the sample z<0 of the pump laser pulses (not presented) co-focused on the surface z=0 with the probe laser pulses. When the pump laser pulses are incident on the sample before the probe laser pulses, then the acousto-optic interaction between the coherent acoustic pulses and the initial probe light field creates the optical polarization sources of the scattered light. The time-domain Brillouin scattering signal is a result of the heterodyne detection of the initially reflected and acoustically scattered probe light collected by the photodetector.

Figure 1 (a) schematically presents the probe light laser beam focused in the transparent medium (in the half space z>0) on the surface (z=0) of the sample (z<0). A superposition of the incident and reflected probe light beams composes the initial light field distribution in the absence of the CAPs.

Both beams at optical frequency $\omega$ satisfy the Helmholtz equation for the electrical field component of the laser radiation ($E(t,x,y,z) \equiv \tilde{E}(t,x,y,z)e^{i\omega t}$)

$$\left[\left(\frac{\partial^2}{\partial x^2} + \frac{\partial^2}{\partial y^2} + \frac{\partial^2}{\partial z^2}\right) + k^2\right]\tilde{E}(x,y,z) = 0, \tag{1}$$

where $k(\omega) \equiv \omega/c$ denotes the optical wave number, $c$ is the speed of light in the medium. The solutions of the wave equation (1) can be obtained via classical decomposition of the three-dimensional wave fields into propagating plane waves $\tilde{\tilde{E}}(k_x, k_y, z)$ with the vector components $k_x, k_y$ and $k_z = \sqrt{k^2 - k_x^2 - k_y^2}$ ($\tilde{E}(x,y,z) = \iint_{-\infty}^{\infty} \tilde{\tilde{E}}(k_x, k_y, z)e^{-i(k_x x + k_y y)}dk_x dk_y$), satisfying the equation

$$\left[\frac{\partial^2}{\partial z^2} + k_z^2\right]\tilde{\tilde{E}}(k_x, k_y, z) = 0. \tag{2}$$

The transverse distribution of the plane wave amplitudes of both incident and scattered light in Fig. 1 (a) are linked at z=0 to the transverse distribution of the incident probe beam $\tilde{\tilde{E}}\underline{}^{probe}(k_x, k_y)$ via $\tilde{\tilde{E}}\underline{}^{incident}(k_x, k_y) \equiv \tilde{\tilde{E}}\underline{}^{probe}(k_x, k_y)$, $\tilde{\tilde{E}}_+^{reflected}(k_x, k_y) = r\tilde{\tilde{E}}\underline{}^{incident}(k_x, k_y) = r\tilde{\tilde{E}}\underline{}^{probe}(k_x, k_y)$, where the reflection coefficient of the probe light at the surface z=0 is denoted by $r$. It is assumed that the probe light does not penetrate in the sample z<0. Thus, the solution of Eq. (2) for the initial probe light field in Fig. 1 is:

$$\tilde{\tilde{E}}^{initial}(k_x, k_y, z) = \tilde{\tilde{E}}\underline{}^{probe}(k_x, k_y)\left(re^{-ik_z z} + e^{ik_z z}\right). \tag{3}$$

When, in addition to ultrashort probe laser pulse, the ultrashort pump laser pulse is focused on the pump light absorbing sample (z<0), the generation of the coherent acoustic waves via optoacoustic conversion takes place [21,22,38]. In Fig. 1 (b) the diffracting strain CAP $\eta(t,x,y,z)$ launched in the medium z>0 is symbolically sketched for a case when the initial radius of the acoustic beam, controlled by the pump laser focusing, is larger than the radius of the probe light focus. The presence of the CAP in the medium induces via the acousto-optic interaction (photoelastic effect) the nonlinear optical polarization and modifies Eq. (1)

$$\left[\left(\frac{\partial^2}{\partial x^2} + \frac{\partial^2}{\partial y^2} + \frac{\partial^2}{\partial z^2}\right) + k^2\right]\tilde{E}(t,x,y,z) = -k^2 n^2 p\eta(t,x,y,z)\tilde{E}(t,x,y,z). \tag{4}$$

In Eq. (4) $n$ and $p$ denote the refractive index and the photoelastic constant of the medium, while $n^2 p \eta(t, x, y, z) \tilde{E}(t, x, y, z) \equiv \tilde{P}(t, x, y, z)$ is the acoustically-induced nonlinear optical polarization [39]. It is worth reminding here that the parametric dependence of the scattering process and on the electric field amplitudes on time via the time dependent strain CAP in Eq. (4) is due to the several orders of magnitude difference between the frequencies of light and acoustic waves. Because the acousto-optic interaction is weak, Eq. (4) can be solved in the single-scattering approximation [40], where the initial "strong" light field creates nonlinear polarization, which emits additional much weaker light waves in the medium

$$\left[\left(\frac{\partial^2}{\partial x^2} + \frac{\partial^2}{\partial y^2} + \frac{\partial^2}{\partial z^2}\right) + k^2\right] \tilde{E}^{scattered}(t, x, y, z) = -k^2 \tilde{P}^{initial}(t, x, y, z). \tag{5}$$

The light waves scattered by the CAP are symbolically presented in Fig. 1 (b). The equation for the transverse distribution of the scattered light, which follows from Eq. (5), is

$$\left[\frac{\partial^2}{\partial z^2} + k_z^2\right] \tilde{\tilde{E}}^{scattered}(t, k_x, k_y, z) = -k^2 \iint_{-\infty}^{\infty} \tilde{P}^{initial}(t, x, y, z) e^{i(k_x x + k_y y)} dx\, dy =$$

$$-k^2 \tilde{\tilde{P}}^{initial}(t, k_x, k_y, z). \tag{6}$$

At this point it is insightful to reveal the relation between the transverse spectra of the polarization $\tilde{\tilde{P}}^{initial}(t, k_x, k_y, z)$, of the initial light field $\tilde{\tilde{E}}^{initial}(\omega, k_x, k_y, z)$ in Eq. (3) and of the acoustic beam. The wide-frequency spectrum acoustic strain beam can be presented as

$$\eta(t, x, y, z) = \frac{1}{(2\pi)^3} \iiint_{-\infty}^{\infty} \tilde{\tilde{\eta}}(\Omega, q_x, q_y) e^{i\Omega t - i(q_x x + q_y y)} e^{-i q_z z} d\Omega\, dq_x\, dq_y, \tag{7}$$

where $q_x, q_y$ and $q_z = \sqrt{q^2 - q_x^2 - q_y^2}$ are the components of the wave vectors $q = \Omega/v$ of the acoustic plane waves composing the beam, $\Omega$ and $v$ are acoustical frequency and velocity, respectively. Note, that in order to capture the basic phenomena contributing to the TDBS it is sufficient to model picosecond CAP with a characteristic duration $\tau_{acoustic} \equiv \tau_a$ as the infinitely short one ($\tilde{\tilde{\eta}}(\Omega, q_x, q_y) = \tilde{\eta}(\Omega) \tilde{\tilde{\eta}}(q_x, q_y), \eta(t) = \delta(t/\tau_a), \tilde{\eta}(\Omega) = \tau_a$) [8,9]. When the plane wave decompositions of the initial light field

$$\tilde{E}^{initial}(x, y, z) = \frac{1}{(2\pi)^2} \iint_{-\infty}^{\infty} \tilde{\tilde{E}}^{probe}(k_x', k_y') e^{-i(k_x' x + k_y' y)} \left(r e^{-i k_z' z} + e^{i k_z' z}\right) dk_x'\, dk_y'$$

and of the acoustic strain beam Eq. (7) are substituted in the polarization $\tilde{P}^{initial}(t, x, y, z)$, the integrations over the spatial coordinates in Eq. (6) reveal the momentum conservation laws between the projections of the light and sound wave vectors on the (x, y) plane

$$\int_{-\infty}^{\infty} e^{i(k_x - k_x' - q_x)x}dx = 2\pi\delta(k_x - k_x' - q_x), \ \int_{-\infty}^{\infty} e^{i(k_y - k_y' + q_y)y}dy = 2\pi\delta(k_y - k_y' - q_y),$$

demonstrating the relation of the polarization and the scattered light angular spectra to the convolution of the angular spectra of the CAP and the initial light

$$\tilde{\tilde{P}}^{initial}(t, k_x, k_y, z) = n^2 p \tau_a \frac{1}{(2\pi)^3} \Big[ \iiint_{-\infty}^{\infty} \tilde{\eta}(q_x, q_y) e^{i\Omega t} e^{-iq_z z} \tilde{\tilde{E}}^{probe}(k_x - q_x, k_y - q_y)$$

$$\left( re^{-ik_z(k,q)z} + e^{ik_z(k,q)z} \right) d\Omega dq_x dq_y \Big], \ \ k_z(k, q) \equiv \sqrt{k^2 - (k_x - q_x)^2 - (k_y - q_y)^2}. \tag{8}$$

The solution of Eq. (6) for the angular amplitude of the scattered light propagating in the positive direction of the z axis, which satisfies, the reflection boundary condition at z=0, is

$$\tilde{\tilde{E}}_+^{scattered}(t, k_x, k_y, z > z_a(t)) = \frac{i}{2k_z}(-k^2)\int_{-\infty}^{\infty} \tilde{\tilde{P}}^{initial}(t, k_x, k_y, z')\left( re^{-ik_z z'} + e^{ik_z z'} \right)dz'. \tag{9}$$

In the derivation of Eq. (9) it was assumed that the CAP is completely launched in the medium z>0, i.e. that at time $t$ of the observation the interface z=0 is not strained, $\eta(t, x, y, z = 0) = 0$, and, thus, $\tilde{P}^{initial}(t, x, y, z \leq 0) = 0$. The scattered field in Eq. (9) is evaluated at distances $z > z_a(t)$, exceeding the distance $z_a(t)$ of the CAP penetration at the observation time $t$, and, thus, $\tilde{P}^{initial}(t, x, y, z > z_a(t)) = 0$. These two conditions provide the explanation for the integration limits in Eq. (9). The second term of the sum in Eq. (9) is due to scattering by the CAP of the initial light in the positive direction of the z axis. The first term, of the sum in Eq. (9) is due to scattering by the CAP of the initial light in the negative direction of the z axis. This scattered light contributes to the total scattered light propagating in the positive direction of the z axis after the reflection at z=0.

The substitution in Eq. (9) of the angular spectrum of the polarization Eq. (8) reveals the momentum conservation law between the projections of the light and sound wave vectors on the z axis:

$$\frac{1}{2\pi}\int_{-\infty}^{\infty}\left\{ e^{i[k_z(k,q)+k_z-q_z]z} + r\left[ e^{i[k_z(k,q)-k_z-q_z]z}e^{-i[k_z(k,q)-k_z+q_z]z} \right] + r^2 e^{-i[k_z(k,q)+k_z+q_z]z}\right\}dz =$$

$$\left\{ \delta[q_z - k_z(k,q) - k_z] + r\left( \delta[q_z - k_z(k,q) + k_z] + \delta\left[ q_z + k_z(k,q) - k_z \right] \right) + r^2\delta\left[ q_z + k_z(k,q) + k_z \right] \right\}$$

The second and the third conservation laws are for the forward light scattering. They require acoustic phonons, propagating at large angles to the probe light axis, which are not available in acoustic beam propagating collinearly to the probe light. Thus, only the first and the forth conservation laws, which are for the backward light scattering, $q_z = \pm[k_z(k,q) + k_z]$, contribute to the solution for the scattered field angular spectrum

$$\tilde{\tilde{\tilde{E}}}_+^{scattered}\left(k_x, k_y, z > z_a(t)\right) = -k^2 n^2 p \tau_a \left(\frac{i}{2k_z}\right) \frac{1}{(2\pi)^3} \int_{-\infty}^{\infty} dz' \iiint_{-\infty}^{\infty} \tilde{\tilde{\eta}}(q_x, q_y) e^{i\Omega t} e^{-iq_z z'}$$

$$\tilde{\tilde{\tilde{E}}}_-^{probe}(k_x - q_x, k_y - q_y)\{e^{i[k_z(k,q)+k_z]z'} + r^2 e^{-i[k_z(k,q)+k_z]z'}\} d\Omega dq_x dq_y. \tag{10}$$

In this solution the first term in the figure brackets is due to the backscattering by the CAP of the probe light presented in Fig. 1 (a) as incident. This process is due to acoustic phonon absorption, which leads to anti-Stocks scattered light. The second term in the figure brackets is due to the backscattering by the CAP of the probe light presented in Fig. 1 (a) as reflected. This process is due to phonon emission. The scattered Stocks light starts to propagate in the positive direction of the z axis towards the photodetector only after the reflection at the interface z=0 [8].

The scattered light is of much smaller amplitude than the reflected light. The information carried by the scattered light is revealed via optical heterodyne detection by mixing the acoustically scattered light with the reflected probe light at the photodetector, which response is linear in absorbed light energy and hence quadratic in the amplitude of the light field. When the photodetector collects all the light from the acoustically scattered and reflected beams, the energy at the photodetector can be found from the angular distributions of the scattered and probe light as follows:

$$W^{detected} = W^{reflected} + W^{heterodyned} + W^{scattered} \sim rr^* \iint_{-\infty}^{\infty} \tilde{\tilde{E}}_-^{probe}(k_x, k_y) \left(\tilde{\tilde{E}}_-^{probe}(k_x, k_y)\right)^* dk_x dk_y \ +$$

$$2Re\left\{\iint_{-\infty}^{\infty} \tilde{\tilde{E}}_+^{scattered}(k_x, k_y) \left(r\tilde{\tilde{E}}_+^{probe}(k_x, k_y)\right)^* dk_x dk_y\right\} + \iint_{-\infty}^{\infty} \tilde{\tilde{E}}_+^{scattered}(k_x, k_y) \left(\tilde{\tilde{E}}_+^{scattered}(k_x, k_y)\right)^* dk_x dk_y$$

Only the second term, providing the frequency mixing of the reflected and scattered light, is relevant to the time-domain Brillouin scattering technique. With the use of Eq. (10) it takes the form

$$W^{heterodyned} \sim k^2 n^2 p \tau_a \frac{1}{(2\pi)^3} Im\left\{\int_{-\infty}^{\infty} dz \iiint_{-\infty}^{\infty} d\Omega dq_x dq_y \tilde{\tilde{\eta}}\left(q_x, q_y\right) e^{i\Omega t} e^{-iq_z z}\right. \tag{11}$$

$$\left. \iint_{-\infty}^{\infty} \frac{1}{k_z} \tilde{\tilde{E}}_-^{probe}(k_x - q_x, k_y - q_y) \left(r\tilde{\tilde{E}}_-^{probe}(k_x, k_y)\right)^* \{e^{i[k_z(k,q)+k_z]z} + r^2 e^{-i[k_z(k,q)+k_z]z}\} dk_x dk_y\right\}.$$

Here and in the following the parameters *n* and *p* of the medium were assumed real for the compactness. Accounting for the complex-valued parameters is straightforward [8]. In Eq. (11) the double integral over all possible probe light wave vector projections on the (x,y) surface, i.e. over the distribution of the plane waves composing the probe light beam, defines the detection sensitivity function $H(k, q, q_x, q_y)$ of the optical heterodyne detection to a particular wave vector components and frequencies of the coherent acoustic beam

$$W^{heterodyned} \sim k^2 n^2 p \tau_a \frac{1}{(2\pi)^3} Im \left\{ \int\limits_{-\infty}^{\infty} dz \iiint\limits_{-\infty}^{\infty} d\Omega dq_x dq_y \tilde{\tilde{\eta}}\left(q_{x'} q_y\right) e^{i\Omega t} e^{-iq_z z} H(k, q, q_x, q_y) \right\}. \quad (12)$$

To get insight in the properties of the sensitivity function it is useful to shift the integration variables $k_x \rightarrow k_x - \frac{q_x}{2}, k_y \rightarrow k_y - \frac{q_y}{2}$.

$$H(k, q, q_x, q_y) = \iint\limits_{-\infty}^{\infty} \frac{1}{\sqrt{k^2 - (k_x - \frac{q_x}{2})^2 - (k_y - \frac{q_y}{2})^2}} \tilde{\tilde{E}}^{probe}\left(k_x - \frac{q_x}{2}, k_y - \frac{q_y}{2}\right) \left(r \tilde{\tilde{E}}^{probe}\left(k_x + \frac{q_x}{2}, k_y + \frac{q_y}{2}\right)\right)^*$$

$$\left\{ e^{i\left[\sqrt{k^2 - (k_x - \frac{q_x}{2})^2 - (k_y - \frac{q_y}{2})^2} + \sqrt{k^2 - (k_x + \frac{q_x}{2})^2 - (k_y + \frac{q_y}{2})^2}\right]z} + r^2 e^{-i\left[\sqrt{k^2 - (k_x - \frac{q_x}{2})^2 - (k_y - \frac{q_y}{2})^2} + \sqrt{k^2 - (k_x + \frac{q_x}{2})^2 - (k_y + \frac{q_y}{2})^2}\right]z} \right\} dk_x dk_y$$

In the paraxial approximation of the diffraction theory [20], the projections of all the wave vectors on the z direction are approximated in the description of the wave amplitude by the full wave vectors, i.e. the transverse wave vector components are neglected. In the description of the phases, the first order corrections, proportional to the square of the transverse and axial components ratio, is taken into account

$$H(q, q_x, q_y) \cong \frac{1}{k} \iint\limits_{-\infty}^{\infty} \tilde{\tilde{E}}^{probe}\left(k_x - \frac{q_x}{2}, k_y - \frac{q_y}{2}\right) \left(r \tilde{\tilde{E}}^{probe}\left(k_x + \frac{q_x}{2}, k_y + \frac{q_y}{2}\right)\right)^*$$

$$\left\{ e^{i2k\left(1 - \frac{k_x^2 + k_y^2}{2k^2} - \frac{q_x^2 + q_y^2}{8k^2}\right)z} + r^2 e^{-i2k\left(1 - \frac{k_x^2 + k_y^2}{2k^2} - \frac{q_x^2 + q_y^2}{8k^2}\right)z} \right\} dk_x dk_y . \quad (13)$$

Important features of the sensitivity function of the heterodyning detection in Eq. (13) are the phases, $\mp i\left(\frac{q_x^2 + q_y^2}{4k}\right)z$, which have the structure of the phases in the paraxial acoustic beams. Note, that they depend only on the optical wave number $k$. They do not depend on the directions of the plane electromagnetic waves in the probe light beam.

The integration over z coordinate in Eq. (12) couples all the projections of the wave vectors on the z axis via the momentum conservation laws

$\delta\left[q_z \mp 2k\left(1 - \frac{k_x^2+k_y^2}{2k^2} - \frac{q_x^2+q_y^2}{8k^2}\right)\right]$. For the paraxial acoustic beam they take the forms $\delta\left[q\left(1 - \frac{q_x^2+q_y^2}{2q^2}\right) \mp 2k\left(1 - \frac{k_x^2+k_y^2}{2k^2} - \frac{q_x^2+q_y^2}{8k^2}\right)\right]$, which demonstrates that the heterodyne detection completely deletes the information on the possible diffraction of the CAP from the time-domain Brillouin scattering signal. The phase factor $-i\left(\frac{q_x^2+q_y^2}{2q}\right)z$ responsible for the diffraction of the acoustic beam (7) is compensated/cancelled by the phase factors $\mp i\left(\frac{q_x^2+q_y^2}{4k}\right)z$ of the heterodyning selectivity function (13) under the backscattering condition $q \cong \pm 2k$. In the paraxial approximation the revealed momentum conservation along the z direction are equivalent to $\delta\left[q \mp 2k\left(1 - \frac{k_x^2+k_y^2}{2k^2}\right)\right] = v\delta\left[\Omega \mp 2kv\left(1 - \frac{k_x^2+k_y^2}{2k^2}\right)\right]$. These momentum conservation laws select the frequencies $\Omega = \pm 2kv \equiv \pm\Omega_B$ in the spectrum of the picosecond acoustic pulse and reconstruct phases $\mp 2kv\left(\frac{k_x^2+k_y^2}{2k^2}\right)t \equiv \mp\left(\frac{k_x^2+k_y^2}{k}\right)z_a(t)$, describing the diffraction of the probe light beam in function of the CAP distance from the focus ($vt \equiv z_a(t)$). The formulated conclusions are confirmed by the following presentation of Eq. (12):

$$W^{heterodyned} \sim kn^2 p\tau_a \frac{1}{(2\pi)^2} Im\left\{\iint_{-\infty}^{\infty} dq_x dq_y \tilde{\tilde{\eta}}(q_x, q_y)\right.$$

$$\left[r^* e^{i\Omega_B t} \iint_{-\infty}^{\infty} \tilde{\tilde{E}}_{-}^{probe}\left(k_x - \frac{q_x}{2}, k_y - \frac{q_y}{2}\right)\left(\tilde{\tilde{E}}_{-}^{probe}\left(k_x + \frac{q_x}{2}, k_y + \frac{q_y}{2}\right)\right)^* e^{-i\left(\frac{k_x^2+k_y^2}{k}\right)z_a(t)} dk_x dk_y \right. +$$

(14)

$$\left.\left. |r|^2 r e^{-i\Omega_B t} \iint_{-\infty}^{\infty} \left(\tilde{\tilde{E}}_{-}^{probe}(k_x - \frac{q_x}{2}, k_y - \frac{q_y}{2})\right)^* \tilde{\tilde{E}}_{-}^{probe}\left(k_x + \frac{q_x}{2}, k_y + \frac{q_y}{2}\right) e^{i\left(\frac{k_x^2+k_y^2}{k}\right)z_a(t)} dk_x dk_y\right]\right\}.$$

The theoretical result in Eq. (14) demonstrates that the compensation of the distance dependent phase of the paraxial CAP, when its motion is monitored by collinear paraxial probe laser pulses via heterodyne interferometry [2,3], is a general phenomenon independent of the transverse distributions of the acoustic and probe laser fields at z=0. For the detection of the CAP motion, the heterodyning creates a complex spectral sensitivity function $H(k, q, q_x, q_y)$ in the wave vectors space. For the backward Brillouin scattering process, the phase of this function contains a part, which is conjugated to the paraxial phase of the CAP and cancels it. Consequently but contra-intuitively, the TDBS signal amplitude variation with time does not depend on the variations of the paraxial CAP amplitude in the diffraction process. The key origin of this phenomenon is the phase sensitive process of the acoustically scattered and the reflected probe light interference.

## 2.2. Gaussian collinear and co-focused acoustic and light beams

For the Gaussian distribution of the probe light field in the focus at z=0, $E(x, y) = E_0 e^{-\frac{x^2+y^2}{a_{probe}^2}}$, the transverse spectrum is also real and Gaussian, $\tilde{E}(k_x, k_y) = E_0(\pi a_{probe}^2)e^{-\left(\frac{a_{probe}}{2}\right)^2(k_x^2+k_y^2)}$. Here the radius of the probe laser beam, $a_{probe}$, is defined at $1/e^2$ level of the light intensity distribution. The spatial decay of the Gaussian laser beam complex amplitude is described in the paraxial approximation by

$$\tilde{E}(x = 0, y = 0, z) = \frac{E_0}{1 \pm i\frac{z}{z_R^o}} = \frac{E_0}{\sqrt{1 + \left(\frac{z}{z_R^o}\right)^2}} e^{\mp i \tan^{-1}\left(\frac{z}{z_R^o}\right)} \ , \qquad (15)$$

where $z_R^o \equiv \pi a_{probe}^2/\lambda_{probe}$ is the Rayleigh range of the probe light beam, while "$\mp$" stand for the light waves propagating in negative and positive directions of the z-axis, respectively. The function $\tan^{-1}(z/z_R^o)$ is the Guoy phase shift. For the Gaussian probe beam the terms in the square bracket of Eq. (14) take the form

$$\frac{2\pi}{a_{probe}^2}\left[E_0(\pi a_{probe}^2)\right]^2 \frac{1}{\sqrt{1 + \left(\frac{z_a(t)}{z_R^o}\right)^2}} \left\{ r^* e^{i\Omega_B t} e^{-i \tan^{-1}\left(\frac{z_a(t)}{z_R^o}\right)} + Rre^{-i\Omega_B t} e^{i \tan^{-1}\left(\frac{z_a(t)}{z_R^o}\right)} \right\} e^{-\frac{1}{2}\left(\frac{a_{probe}}{2}\right)^2(q_x^2+q_y^2)}. \quad (16)$$

In accordance with Eqs. (14) and (16), in addition to above discussed cancellation of the paraxial phase of the CAP beam, the heterodyne detection introduces a transverse filter, $\sim e^{-\frac{1}{2}\left(\frac{a_{probe}}{2}\right)^2(q_x^2+q_y^2)}$, for the acoustic rays composing the CAP. The comparison of the time-dependent part in Eq. (16) with Eq. (15) reveals the relation of the former to the spatial distribution of the complex amplitude in the initial probe light field (Fig. 1 (a)). The dependencies on $z_a(t)$ of the first and the second terms of the sum in Eq. (16) replicate the variations with z of the complex amplitudes in the incident and reflected probe light beams, respectively.

In case of the pump laser pulses absorption at the surface z=0, the distribution of the acoustic strain at the surface is controlled by the power distribution in the Gaussian pump laser focus of radius $a_{pump}$ [21,22], $I_{pump}(x,y) = \frac{2P_{pump}}{\pi a_{pump}^2} e^{-2\left(\frac{x^2+y^2}{a_{pump}^2}\right)}$. Correspondingly, the transverse spectrum of the strain CAP is also Gaussian

$$\tilde{\tilde{\eta}}(q_x, q_y) = \eta_0 \iint_{-\infty}^{\infty} e^{-2\left(\frac{x^2+y^2}{a_{pump}^2}\right)} e^{i(q_x x + q_y y)} dx dy = \eta_0 \frac{(\pi a_{pump}^2)}{2} e^{-\frac{1}{2}\left(\frac{a_{pump}}{2}\right)^2 (q_x^2 + q_y^2)}, \quad (17)$$

where acoustic strain amplitude $\eta_0$ is related to the pump laser power via the coefficient of optoacoustic conversion $C_{oa}$, $\eta_0 = C_{oa} \frac{2P_{pump}}{\pi a_{pump}^2}$. The acoustic beam in real space is also Gaussian with a complex amplitude decaying in the positive direction of z axis following the law

$$\eta(x = 0, y = 0, z) = \frac{\eta_0}{1 - i\frac{z}{z_R^a}} = \frac{\eta_0}{\sqrt{1 + \left(\frac{z}{z_R^a}\right)^2}} e^{i \tan^{-1}\left(\frac{z}{z_R^a}\right)}, \quad (18)$$

where $z_R^a \equiv \frac{\pi a_{pump}^2}{2\lambda_{acoustic}}$ is the Rayleigh range for the paraxial Gaussian beam and $\lambda_{acoustic}$ denotes the wavelength of the acoustic wave.

It is worth noting that in the backward Brillouin scattering, under the current analysis, the angular acoustic frequency is $\Omega = \Omega_B = 2kv = 2\left(\frac{2\pi}{\lambda_{probe}}\right)v$, while $\lambda_{acoustic} = \lambda_{probe}/2$. Consequently, the ratio of the Rayleigh ranges of probe light and coherent acoustic beam does not depend on the wavelengths. It is controlled by the ratio of the radii of the probe and pump laser beams, defined at the same level, $z_R^0/z_R^a \equiv a_{probe}^2/a_{pump}^2$. For the Gaussian acoustic beam the transverse spectrum filtering, imposed by the heterodyne detection of the TDBS signal by Gaussian probe light, is described by

$$\iint_{-\infty}^{\infty} \tilde{\tilde{\eta}}(q_x, q_y) e^{-\frac{1}{2}\left(\frac{a_{probe}}{2}\right)^2 (q_x^2 + q_y^2)} dq_x dq_y = \eta_0 \frac{(\pi a_{pump}^2)}{2} \frac{8\pi}{\left[a_{pump}^2 + a_{probe}^2\right]}.$$

For the reasons discussed in the previous section, the information on the decay of acoustic beam, revealed in Eq. (18) is absent in the TDBS signal. Substituting all the above-derived results for the Gaussian beams in Eq. (14), the TDBS signal can be presented in the normalized form

$$\frac{W^{heterodyned}}{W^{reflected}} = \frac{4|r|(1-R)}{R} \left(\frac{l_a}{\lambda_{probe}}\right) \frac{n^2 p C_{oa} P_{pump}}{\left[a_{pump}^2 + a_{probe}^2\right]} \frac{1}{\sqrt{1 + \left(\frac{z_a(t)}{z_R^0}\right)^2}} \sin\left[\Omega_B t - \varphi - \tan^{-1}\left(\frac{z_a(t)}{z_R^0}\right)\right]. \quad (19)$$

In the description (19) of the so-called Brillouin oscillation, $l_a \equiv v\tau_a$ denotes the characteristic length of the picosecond CAP, $\varphi$ is the phase of the probe light reflection coefficient at z=0 ($r = |r|e^{i\varphi}$, $rr^* = |r|^2 \equiv R$). Note, that Eq. (19) for the diffracting light and sound, reproduces for the acoustic pulse travelling inside the Rayleigh range, i.e. near field, of the probe light, $z_a(t) \ll z_R^o$, the theoretical predictions for the plane waves from Ref. [8], in particular $P^{heterodyned} \sim (1 - R)$. The amplitude of the theoretically predicted Brillouin oscillation depends on the optical Rayleigh range and does not depend on the acoustical one. At the same time, it is not that cumbersome to prove analytically that in the Gaussian beams both the amplitude the acoustically scattered probe light and the energy in the scattered beam depend on both optical and acoustical Rayleigh ranges and are sensitive to the diffraction of the CAP. For example

$$\left| \tilde{E}^{scattered}(x=0, y=0, t) \right| \sim \frac{1}{\sqrt{1 + \frac{1}{4}\left(\frac{z_a(t)}{z_R^o + (z_R^a/2)}\right)^2} \sqrt{1 + \left(\frac{a_{probe}}{a_{pump}} + \frac{a_{pump}}{a_{probe}}\right)^4 \left(\frac{z_a(t)}{z_R^o + (z_R^a/2)}\right)^2}}$$

However, the TDBS signal amplitude obtained by the interferometric heterodyning is not proportional to the amplitude of the scattered light. In accordance to Eq. (19), the diffraction of the CAP does not influence the TDBS signal decay. Thus, it does not shorten the depth of the TDBS imaging and does not broaden the Brillouin line and diminish accuracy of the frequency measurements by the TDBS microscopy [32,33].

### 2.3. Gaussian collinear paraxial acoustic and light beams with shifted foci

The above presented theory was developed for co-focused probe light and CAP beams (see Fig. 1) just to avoid too lengthy formulas. The major result, on the independence of the TDBS signal on the evolution of the CAP amplitude caused by the acoustical diffraction, is valid as well in the case of the probe light and CAP beams focused in different spatial positions along the z axis. In Fig. 2 the TDBS geometry, relevant to imaging inside the semi-infinite medium, z<0, by the probe light and CAP beams launched from the surface z=0, is schematically presented.

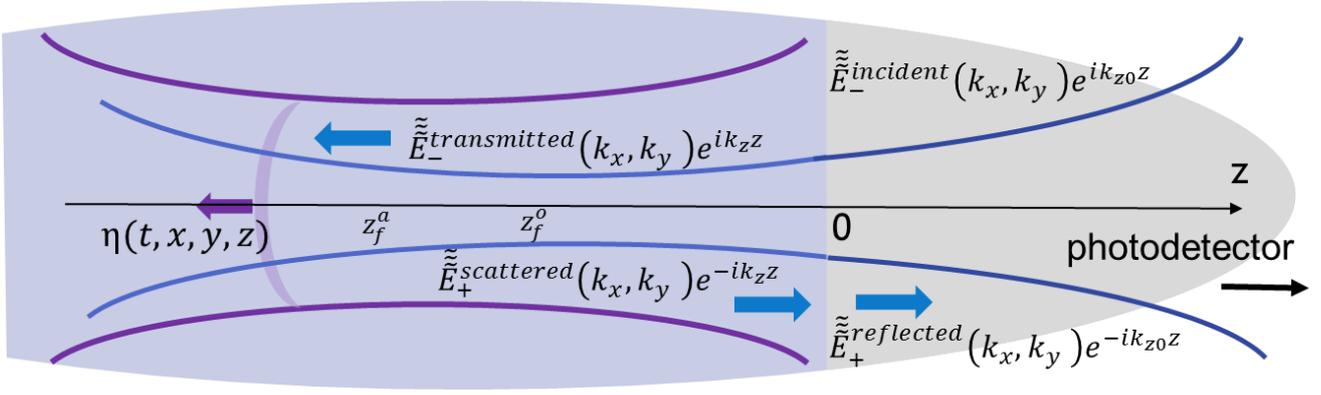

**Figure 2.** **Schematic presentation of probe light and coherent sound fields contributing to time-domain Brillouin scattering signal detection when they are focused at different depths $z_f^o \neq z_f^a$ inside the medium.** Initial probe light field, that could be created in the system in the absence of the coherent acoustic field, is due to the laser pulses incident from the medium z>0 on the sample surface z=0 and transmitted into the TDBS tested medium z<0, which is transparent for the probe light. Coherent acoustic pulse $\eta(t, x, y, z)$ is launched in the tested media z<0 due to the absorption there of the pump laser pulses (not presented) co-focused on the surface z=0 with the probe laser pulses. When the pump laser pulses are incident on the sample before the probe laser pulses, then the acousto-optic interaction between the coherent acoustic pulses and the initial probe light field creates the optical polarization sources of the scattered light. The time-domain Brillouin scattering signal is a result of the heterodyne detection of the initially reflected and acoustically scattered probe light collected by the photodetector.

A part of the incident probe light transmitted across the interface of the first medium (z<0) and the second medium (z>0), propagating in the negative direction of the z axis, creates the initial light beam for monitoring the CAP. The probe light is scattered by the CAP only in the positive direction of z axis. This backward scattered Stokes light beam after transmission across the interface z=0 is heterodyned at the photodetector by a part of the incident probe light reflected at the interface (see Fig. 2). From the mathematical analysis point of view, the only significant difference of the schemes presented in Fig. 1 and Fig. 2 consists in the different positions of the probe light and CAP beams foci on the z axis, $z_f^o \neq z_f^a$ (see Fig. 2). The related modifications of the theoretical formulas of Sections II.1 and II.2 are reduced to decomposition of the acoustic and transmitted light beams into plane waves $e^{i\Omega t - i(q_x x + q_y y)} e^{iq_z(z - z_f^a)}$ and, respectively, $e^{i\omega t - i(k_x' x + k_y' y)} e^{ik_z'(z - z_f^o)}$, with different foci. The most important differences of the BO predicted for the scheme in Fig. 2 in comparison with the BO described by Eq. (19) for the scheme in Fig. 1, are the following. In the former there is only one Brillouin scattering process, because the initial light beam in the tested medium is unidirectional. Thus, the disappearance of the Brillouin oscillation due to the destructive interference of two scattering processes, described by the factor $(1 - R)$ in Eq. (19), is impossible. The dependence of the amplitude of the Brillouin oscillation on time replicates the amplitude distribution along the z axis in the probe light beam in the medium ($\sim 1/\sqrt{1 + \left(\frac{z_a(t) - z_f^o}{z_R^o}\right)^2}$). The Guoy phase shift in the scheme of Fig. 2 takes the form $\tan^{-1}\left(\frac{z_a(t)-z_f^o}{z_R^o}\right)$, respectively. The acoustic waves filtering amplitude factor $e^{-\frac{1}{2}\left(\frac{a_{probe}}{2}\right)^2\left(q_x^2+q_y^2\right)}$, revealed in Eq. (16), acquires in the scheme of Fig. 2 the additional phase multiplier $e^{-i\frac{(z_f^a-z_f^o)}{4k}\left(q_x^2+q_y^2\right)}$. This results in the additional amplitude factor, $\frac{1}{\sqrt{1+\left(\frac{z_f^o-z_f^a}{z_R^o+z_R^a}\right)^2}}$, and the additional phase shift, $\sim\tan^{-1}\left(\frac{z_f^o-z_f^a}{z_R^o+z_R^a}\right)$, in the BO. Although the last two factors depend of the acoustical Rayleigh range, they are stationary factors controlled by the constant distance between the optical probe and the acoustical foci. Consequently, the TDBS signal in scheme of Fig. 2 is independent of the focusing/diffraction of the acoustic beam, as it could be expected from the analysis of the fundamental Brillouin scattering and heterodyning processes discussed earlier in this section. The dynamics of the TDBS signal depends only the focusing/diffraction of the probe light pulses.

## 3. Discussion

### 3.1. Interpretation of the existing experimental observations

The theoretical predictions presented in Section II can be compared with the experimental results presented in Ref. [23]. In Ref. [23] the CAPs were sharply focused on the axis of the sub-micrometer diameter optical fiber while the probe light field was a weakly diffracting beam with a focus approximately at a half distance between the cylindrical opto-acoustic transducer at the fiber surface and the fiber axis (see Figs. 1 (a) and 7 in ref. 23). However, the variations of the experimental TDBS signal [23] revealed no sign of the CAP focusing temporal dynamics, in accord with the above developed theoretical predictions.

The developed theory can be applied for the analysis of the experiments [15] where the influence of the CAP diffraction on the attenuation of the BO amplitude was hypothesized. In Ref. [15] the CAPs were launched in a spherical microcapsule of the 4 **μm** radius, composed of a glassy polymer shell encapsulating a liquid perfluorinated core. The capsule was deposited on the metallic substrate and both pump and probe laser pulses were focused through the capsule on the contact between the capsule and the metal. The radius of the probe light beam was 1 **μm**, while the radius of the pump was 2 **μm**. However, the radius of the acoustic beam emitted in the capsule was controlled not by the pump laser focus but by much smaller radius of a contact of the spherical capsule with the substrate, which was estimated as $a_{contact} \approx 0.3$ **μm**. Note, that the contact radius is also significantly smaller than the radius of the probe laser focus. Thus, based on the estimates of the acoustic Rayleigh ranges, it was hypothesized [15] that the deviation of the BO decay law from the exponential one, related to acoustic absorption, is due to the transition of the CAP from the near to far field of the acoustic source. However, for the 0.3 **μm** radius of the source and the experimental acoustic Brillouin wavelength of about 0.15 **μm**, the half angle of the acoustic beam divergence [19] is estimated as $\frac{\lambda_a}{\pi a_{contact}} \approx 0.16$ radians. So, the acoustic beam is paraxial and, in accordance with above developed theory, its diffraction should not influence the dynamics of the TDBS signal amplitude. We hypothesize here that either in Ref. [15] the radius of the contact was overestimated and the propagation of CAP significantly deviates from paraxial one, or the contact acts as an aperture on the probe light reducing the effective radius of the initial probe light beam involved in the TDBS. In the latter case, the experimental observations [15] could be caused by the significant diminishing of the Rayleigh range for the probe light beam reflected from the contact surface in comparison to the incident probe light. Of cause, potentially, the overestimations of both the acoustic and probe beam effective radii are possible.

The authors of Refs. [32] and [33] discussed the roles of both acoustical and probe light diffraction in the experimentally observed non-exponential decay of the TDBS signals detected at different NAs of the microscope objectives. The NAs were varying from 0.1 up to 1.3. The interpretation of the experimental signals [33] is based on the assumption that the TDBS signal is proportional to the product of the local amplitudes of the paraxial Gaussian acoustical and probe light beams, suggesting that both these diffraction phenomena should be accounted for the correct extraction of the sound absorption coefficient from the TDBS signal decay. However, the analytical theory presented in the Section II suggests that the applied intuitive assumption is not valid. The estimates reveal that the photo-generated acoustic beams [32,33] are paraxial, $\frac{\lambda_a}{\pi a_{pump}} \leq 0.15$, for all used NAs. The analytical theory predicts that such acoustic beams are not influencing the temporal dynamics of the TDBS amplitude, at least when probing their propagation by collinear paraxial probe light beams. In the case of the diffraction-limited focusing of light, the half angle of the probe light beam divergence in glass, $n \approx 1.46$, is estimated as $\frac{\lambda_{probe}}{\pi a_{probe}} \approx 0.3 \cdot NA$ . So, in the experiments [32,33] the influence of the acoustic diffraction on the TDBS amplitudes dynamics could be expected only for $NA \geq 0.5$ due to stronger diffraction of probe light beam in comparison with sound beam, and the deviation of the probe light rays from paraxial propagation. However, this expectable influence is weak in comparison with the direct influence on the TDBS amplitude of the inhomogeneous spatial distribution of the diffracting probe light. Note, that the paraxial approximation holds very well for the acoustic beam generated by the pump light but could become less accurate for probe light because in the experiments [32,33] the radius of the pump laser beam is larger than the radius of probe laser beam.

## 3.2. On the relations between the TDBS and SBS

In view of the knowledge that coherent acoustic waves are involved in the stimulated Brillouin scattering of light [34,35] and that the phenomenon of the optical wave front phase conjugation [35-37] is one of the fingerprints of the SBS, it is worth to compare qualitatively the TDBS to the SBS. From the physics point of view, these processes are principally different.

First, although the phonons initiating the TDBS are coherent, they are launched in the medium by additional external means. They are not generated by the counter-propagating incident and backscattered light as it takes place in SBS. Although, the stimulated emission of phonons in the Stocks backward Brillouin scattering process takes place, the contribution of these phonons to further scattering of the probe light in the TDBS is negligible as they are not further amplified and their amplitudes are much smaller in comparison with amplitudes of the initially launched phonons. In the SBS the coherent acoustic waves are generated due to the interaction of the coherent incident light beam with the dominant coherent part of the scattered light beam via one of the possible physical mechanisms, i.e. the electrostriction, for example [21,22,38]. Saying differently, the coherent acoustic phonon in the SBS is the result of the selective amplification of a particular thermal phonon via the processes of the stimulated phonon emission. The amplitude of these phonons in SBS largely exceeds the amplitudes of the thermal phonons initially existing in the medium. The amplitude of the coherent phonons generated by the interaction of the light fields, for example via the electrostriction or other processes [21,38], is completely negligible in TDBS in comparison with the amplitudes of the externally launched coherent phonons.

Second, although the phonons initiating the TDBS are coherent, the probe light scattered by them is not phase conjugated to the incident light and contains the information on the phase evolution in the launched coherent phonon field. It is the heterodyne detection of the scattered and reflected light, which makes the TDBS signal insensitive to the wave front of the phonon in the launched coherent acoustic beam. In the SBS the front of the coherent acoustic phonon is created in the process of the phonon generation by the counter-propagating strong light fields. Saying differently, the coherent phonon wave front is selected in the SBS by the process of the thermal phonons selective stimulated amplification. This acoustic front optimally fits the momentum conservation laws for the photon-phonon interactions and induces wave front reversal of the scattered light. It could be expected that the physical processes in the TDBS would approach those in the SBS with the increase in the intensity of the probe light and/or of the initial coherent acoustic field when the phonons generated/amplified by the counter-propagating light fields will dominate in the coherent acoustic field over the initially launched.

## 4. Conclusions

The developed theory of the TDBS in collinear paraxial acoustical and optical beams predicts that, contra-intuitively, the TDBS amplitude dynamics does not depend on the variations of the coherent acoustic pulse amplitude that could be caused by the diffraction/focusing of the CAPs. This theoretical prediction correlates with earlier reported experimental observations. The theory provides the explanations to some existing experimental observations, which are different from the earlier suggested. Finally, the developed theory provides insight in the possible ways for the optimization of the TDBS based imaging and spectroscopy. It predicts that, as far as the CAPs, photo-generated by the pump laser pulses, and probe laser beam are paraxial the lateral resolution of the imaging could be enhanced by sharper focusing of the pump laser beam without diminishing the imaging depth and the spectral resolution. The developed theory reveals earlier unexpected features in the TDBS scattering of the paraxial coherent sound and probe light beams that could be useful and advantageous in the existing and future applications of the TDBS imaging and microscopy technique for the fundamental and applied research in a variety brunches of science.

### Acknowledgements

This work was supported by the French National Research Agency (ANR, France) through the grant <ANR-18-CE42-017>. The author thanks T. Dehoux, O. B. Wright, T. Thréard and S. Raetz for collaboration and discussions.